\documentclass[twocolumn,showpacs,preprintnumbers,amsmath,amssymb,floatfix]{revtex4}

\usepackage{graphicx}
\usepackage{dcolumn}
\usepackage{bm}
\usepackage{fancybox}

\newcommand{\bfk}{{\bf k}}
\newcommand{\bfr}{{\bf r}}

\newcommand{\ocite}[1]{\cite{#1}}

\def\qsgw{QS{\em GW}}

\def\ekn{{\varepsilon_{{\bf k}n}}}

\def\Psikn{\Psi_{{\bf k}n}}

\def\Psikdnd{\Psi_{{\bf k}'n'}}

\def\H0{H^0}

\newcommand{\req}[1]{\mbox{Eq.~(\ref{#1})}}

\def\ekn{{\varepsilon_{{\bf k}n}}}

\def\Psikn{\Psi_{{\bf k}n}}

\def\connect#1{\leavevmode{\setbox1=\hbox{#1}\copy1%
\raise .2\ht1 \vbox{\moveleft \wd1\vbox{\hrule width \wd1 height .5pt depth 0pt}}%
}}

\def\ftn[#1]{\rlap{\footnotemark[#1]}}

\begin{document}

\title{Impact ionization rates for Si, GaAs, InAs, ZnS, and GaN 
in the $GW$ approximation}
\author{Takao Kotani}
\affiliation{School of Materials, Arizona State University, Tempe, AZ, 85284}
\author{Mark van Schilfgaarde}
\affiliation{School of Materials, Arizona State University, Tempe, AZ, 85284}
\date{\today}

\begin{abstract}
We present first-principles calculations 
of the impact ionization rate (IIR) 
in the $GW$ approximation ($GW$A) for semiconductors.
The IIR is calculated from the quasiparticle (QP) width 
in the $GW$A, since it can be identified as the
decay rate of a QP into lower energy QP plus an 
independent electron-hole pair. 
The quasiparticle self-consistent $GW$ method was used to 
generate the noninteracting hamiltonian the $GW$A requires as input.
Small empirical corrections were added so as to reproduce experimental band gaps.
Our results are in reasonable agreement with previous work, 
though we observe some discrepancy.  In particular we find
high IIR at low energy in the narrow gap semiconductor InAs.
\end{abstract}


\maketitle
\section{introduction}
The electron-initiated impact ionization
is a fundamental process in 
semiconductors where a high energy electron decays 
into an another low-energy electron together 
with an electron-hole pair \cite{ferry00}.
The impact ionization rate (IIR), which originates from the coulomb interaction between electrons, 
is a critical factor affecting transport 
under high electric field, as described by the
Boltzmann transport equation (BTE). 
It is important in narrow gap semiconductors, especially for 
ultrasmall devices.
Impact ionization is also used in
avalanche photodiodes, and to supply electron-hole pairs
for electroluminescence. Recently it has stimulated interest as a
mechanism to improve efficiency in photovoltaic devices\cite{klimov04}.

The IIR has been calculated with empirical pseudopotentials (EPP) 
in order to include realistic energy bands
\cite{kane67,sano92,sano94,sano95,jung96}.
Sano and Yoshii calculated the IIR for Si \cite{sano92,sano94}
and obtained reasonable agreement with experimental data.
They also studied other materials \cite{sano95}, treating the
transition matrix element $M$
as a parameter (constant matrix approximation).
Jung et al. \cite{jung96} used an EPP to calculate the IIR in GaAs.
They calculated $M$ 
including explicit calculation of the dielectric function $\epsilon({\bf q},\omega)$,
rather than assuming a model form.

Recently, two groups have calculated the IRR using the density-functional formalism 
to generate the one-body eigenfunctions and energy bands.
Because the standard local density approximation (LDA) underestimates semiconductor bandgaps
while the IIR is very sensitive to this quantity,
the standard LDA is not suitable.
Picozzi et al. used a screened-exchange generalization\cite{Seidl96} of the LDA
\cite{picozzi02L,picozzi02}, 
and Kuligk et al. employed the exact exchange\cite{Kotani94,Stadele97} formalism \cite{kuligk05}.
Both groups used model dielectric functions for the
dynamically screened coulomb interaction $W(\bfr,\bfr',\omega)$.

Here we will present (nearly) {\it ab initio}
calculations of the IIR without model assumptions.
First, our noninteracting hamiltonian $\H0$ is generated within
quasiparticle self-consistent $GW$ (\qsgw) formalism.
We have shown that \qsgw\ works very well
for wide range of materials 
\cite{Faleev04,vans06,kotani07a,kotani07b,chantis07f}. 
Because the IIR is highly sensitive to the bandgap,
we add a small empirical scaling of the exchange-correlation potential
so as to reproduce the experimental fundamental gap $E_{G}$.
Corrections for semiconductors 
are small and systematic as shown below.
Second, $W$ is calculated from the \qsgw\ noninteracting hamiltonian. 
The IIR is identified with the decay rate (or linewidth) of the
quasiparticle (QP), which is
calculated from the imaginary part of the self-energy, as we describe below.
Our method thus contains only one parameter, to correct the band gap. 
As we have shown\cite{chantis06a}, this parameter is small and is approximately independent of material.
In principle our method can predict the IIR in unknown systems, 
and also for inhomogeneous systems such as grain 
boundaries, quantum dots, or impurities, where the IIR should 
be strongly enhanced because momentum conservation is much more 
easily satisfied.  Thus the present \emph{ab initio} method should be superior 
to prior approaches.
Applications to such systems will be useful in devices that need
to suppress or enhance electron-hole pair-generation
from impact ionization.

After a theoretical discussion, we present some results.
They are in reasonable agreement with 
previous calculations, except for InAs where 
IIR is calculated to be much higher than what Sano and Yoshii found \cite{sano95}.

\section{Method}

The first step is to determine a good one-body Hamiltonian $\H0$ 
which describes QPs.
We obtain $\H0$ from \qsgw\ calculations\cite{Faleev04,vans06,kotani07a}. 
As we explain in Sec.\ref{sec:res}, we follow Ref.\cite{chantis06a}, 
and modify $\H0$ by a simple empirical scaling
($\alpha$-correction) to ensure the fundamental gap reproduces experiment.
From this modified $\H0$ we obtain a set of eigenvalues 
$\{\varepsilon_{{\bfk}n}\}$ and eigenfunctions $\{\Psikn\}$,
which are used to calculate
the self-energy $\Sigma(\bfr,\bfr',\omega)$
within the $GW$A, $\Sigma= i G \times W$.
The inverse of the QP lifetime $\tau_{\bfk n}^{-1}$
is obtained from the imaginary part of $\Sigma$ as
\begin{eqnarray}
\tau_{\bfk n}^{-1}= 
\frac{2 Z_{\bfk n}}{\hbar} \left|{\rm Im}\Sigma_{\bfk n}\right|,
\label{eq:invt}
\end{eqnarray}
where
${\rm Im}\Sigma_{\bfk n} =
\langle \Psikn |{\rm Im}\Sigma(\ekn) |\Psikn \rangle 
= \int d^3r \int d^3r' \Psikn^*(\bfr)
{\rm Im}\Sigma(\bfr,\bfr',\ekn) \Psikn(\bfr')$.
By
${\rm Im}\Sigma(\ekn)$ we mean the anti-hermitian part.
$Z_{\bfk n}$ is the wave function renormalization factor
to represent the QP weight.
${\bfk}$ denotes the wave vector in the first Brillouin zone (BZ),
and $n$ the band index.
The expression Eq.\ref{eq:invt} for $\tau_{\bfk n}^{-1}$ 
is derived in Appendix B of Ref.\cite{echenique00}.
${\rm Im}\Sigma$ is obtained from the 
the imaginary part of the convolution of
$G$ and $W$.
For an unoccupied state ${\bf k} n$, it is
\begin{eqnarray}
{\rm Im}\Sigma_{{\bf k} n} 
&=&-\int d^3r d^3r' \sum_{\bfk' n'} 
\Psikn^*(\bfr) \Psikdnd(\bfr) 
\Psikdnd^*(\bfr') \Psikn(\bfr') \nonumber \\
&\times& \pi {\rm Im}W(\bfr,\bfr',\ekn-\varepsilon_{\bfk' n'}),
\label{eq:imsig}
\end{eqnarray}
where states ${\bf k}'n'$ are restricted to those
for which $\varepsilon_{\rm F} <\varepsilon_{\bfk' n'} 
<\varepsilon_{\bfk n}$.
$W$ is calculated in the random 
phase approximation (RPA) as
\begin{eqnarray}
W= v+ v \chi v = (1-v \chi_0)^{-1} v,
\label{eq:w}
\end{eqnarray}
where $v$ is the coulomb interaction; 
$\chi$ is the full polarization function in the RPA,
and $\chi_0$ is the non-interacting polarization function.
With Eqs.(\ref{eq:invt},\ref{eq:imsig},\ref{eq:w}),
$\tau_{\bfk n}^{-1}$ is calculated from $\H0$ in principle.
In the Lehmann representation, $\chi$ is 
\begin{eqnarray}
\chi(\bfr,\bfr',\omega)= 
\sum_m \langle 0|\hat{n}(\bfr) |m \rangle 
             \langle m|\hat{n}(\bfr') |0 \rangle \nonumber \\
    \times   \left(\frac{1}{\omega - \omega_m - i\delta}
            -\frac{1}{\omega + \omega_m + i\delta} \right),
\label{eq:lehmannw}
\end{eqnarray}
where $|m \rangle$ denotes the eigenstates (intermediate states)
with excitation energy $\omega_m$ relative to
the ground state $|0 \rangle$.
Here $\hat{n}(\bfr)$ is the density operator.

In the RPA, $|m \rangle$ are
the eigenfunctions of a two-body (one electron and one hole)
eigenvalue problem in the RPA.
In simple cases such as the homogeneous electron gas,
$|m \rangle$ for high $\omega_m$ are identified as
plasmons; $|m \rangle$ for low $\omega_m$ are as
independent motions of an electron and a hole.  Thus
$\tau_{\bfk n}^{-1}$ for low energy electrons
calculated in $GW$A can be identified as 
the transition probability to such states 
for the independent motion of an electron and a hole 
together with an electron; that is, we identify $\tau_{\bfk n}^{-1}$ as the IIR.

There are some questionable points for the identification.
It might be not so easy in some cases to identify a state $|m \rangle$ 
as such a independent motion because the electron-hole pair 
can be hybridized with plasmons.
However, such hybridization is sufficiently small for the simple 
semiconductors treated here, because plasmons appear
only at high energies as $\omega_m\gtrsim$ 1 Ry.
Another problem is that the final state consisting of
two electrons and one hole is not symmetrized for the 
electrons in the $GW$A. Thus Fermi statistics are not satisfied.
Below we discuss how much error it causes.

Our formula \req{eq:invt} for $\tau_{\bfk n}^{-1}$ 
is different from the customary expression found in the 
literature
\cite{kane67,sano92,sano94,sano95,jung96}, e.g,
see Eq.(1) in Ref.\cite{kane67}.
It is written as
\begin{eqnarray}
\tau_{\bfk n}^{-1} &=& \frac{4 \pi}{\hbar}
\sum_{\bfk' n'}\sum_{\bfk_1 n_1}\sum_{\bfk_2 n_2} |M|^2 \nonumber \\
&&\times \delta(\ekn - \varepsilon_{\bfk' n'}- \varepsilon_{\bfk_1 n_1}
      + \varepsilon_{\bfk_2 n_2}),
\label{eq:tau2}
\end{eqnarray}
where $|M|^2 = \frac{1}{2}(|M_{\rm D}|^2 + |M_{\rm E}|^2 + |M_{\rm D}-M_{\rm E}|^2)$
includes both direct and exchange processes.
The sum over $\bfk',\bfk_1,\bfk_2$ is restricted to satisfy $\bfk=\bfk'+\bfk_1-\bfk_2$.
The matrix element $M_{\rm D}$ for the direct process is
\begin{eqnarray}
M_{\rm D}&=&\Bigg| \int d^3r d^3r''  
\Psikn(\bfr) \Psikdnd^*(\bfr) \times \nonumber \\
&&W(\bfr,\bfr'',\ekn - \varepsilon_{\bfk' n'})
\Psi_{{\bf k}_1 n_1}^*(\bfr'') \Psi_{{\bf k}_2 n_2}(\bfr'') \Bigg|. 
\label{eq:md}
\end{eqnarray}
$M_{\rm E}$ for the exchange process is the same as $M_{\rm D}$, 
except that the two electrons in final states 
($\bfk'n' \leftrightarrow \bfk_1 n_1$) are exchanged.
\req{eq:tau2} can be derived in time-dependent perturbation theory, 
where the final states consists of two electrons and one hole.
This is based on the physical picture that $W$
causes transitions between the Fock states 
made of QPs. However, the final 
states made of the three QPs are interacting each other.
Thus such a picture do not necessarily well-defined. 
This is related to a fundamental problem about 
how to mimic the quantum theory by the BTE. 
Definition of the IIR suitable for the BTE is somehow ambiguous. 
The difference between \req{eq:invt} and \req{eq:tau2}
is related to the ambiguity. 
One is not necessarily better than the other.

%
%
%

To compare \req{eq:invt} with \req{eq:tau2},
let us assume that Im$\chi_0$ is small enough.
Then we have 
\begin{eqnarray}
{\rm Im}W\approx W_{\rm R} \ {\rm Im}\chi_0 \ W_{\rm R} 
\label{eq:imw}
\end{eqnarray}
from \req{eq:w}, 
where $W_{\rm R}= (1-v {\rm Re}\chi_0)^{-1} v$. 
Re$\chi_0$ denotes the hermitian (real) part of $\chi_0$.
If we apply \req{eq:imw} to \req{eq:imsig}, \req{eq:invt} is reduced
to an expression similar to \req{eq:tau2}:
\begin{eqnarray}
\tau_{\bfk n}^{-1} &\approx& \frac{4 \pi Z_{\bfk n}}{\hbar}
\sum_{\bfk' n'}\sum_{\bfk_1 n_1}\sum_{\bfk_2 n_2} |M_{\rm D}|^2 \nonumber \\
&&\times \delta(\ekn - \varepsilon_{\bfk' n'}- \varepsilon_{\bfk_1 n_1}
      + \varepsilon_{\bfk_2 n_2}),
\label{eq:tau1}
\end{eqnarray}
where $M_{\rm D}$ is defined in \req{eq:md}
but with $W_{\rm R}$ instead of $W$.
Through \req{eq:tau1} we can elucidate
the differences between \req{eq:invt} and \req{eq:tau2} 
as follows:
\begin{itemize}
\item[(a)] \req{eq:tau1} (and thus \req{eq:invt}) contain the $Z$ factor.
This is because \req{eq:tau2} was derived without
taking into account the modification of QPs 
by the coulomb interaction.
Typically $Z_{\bfk n}$ is $\sim 0.8$.
\item[(b)] \req{eq:tau1} and \req{eq:invt}
do not include $M_{\rm E}$ contributions.
In the extreme case when $M_{\rm E} = \frac{1}{2} M_{\rm D}$ ,
$|M|^2=0.75 \times |M_{\rm D}|^2$.
This occurs in the Hubbard model when $W$ is a point interaction;
the Feynman diagrams for $M_{\rm D}$ and $M_{\rm E}$
become the same except for their sign.
Theoretically, including $M_{\rm E}$ is advantageous
because it symmetrizes the two electrons in the final state
(though only for ${\rm Im}\chi_0$ in the linear response regime).
Fermi statistics are not perfectly satisfied 
because not all the exchange-pair diagrams are included.
Omitting the exchange contribution reduces the IIR 
by a factor 0.75 at most, as explained above.
\item[(c)] \req{eq:tau1} contains only the real part of $W$,
in contrast to \req{eq:tau2}.
The difference originates from higher order
contributions to ${\rm Im}\chi_0$.
Moreover, when \req{eq:imw} is not satisfied there are further
higher-order contributions to ${\rm Im}\chi_0$.
\end{itemize}
We may have to pay attention to these differences.
For small ${\rm Im}\chi_0$, 
(a) and (b) predominate, and the difference between \req{eq:invt} and 
the Kane formula \req{eq:tau2} should be a factor in the range $0.5$ to 1.
However, this difference is relatively minor on the log scale in the Figure.

\section{results}
\label{sec:res}
\begin{figure*}[htbp]
\centering
\includegraphics[angle=0,scale=.6]{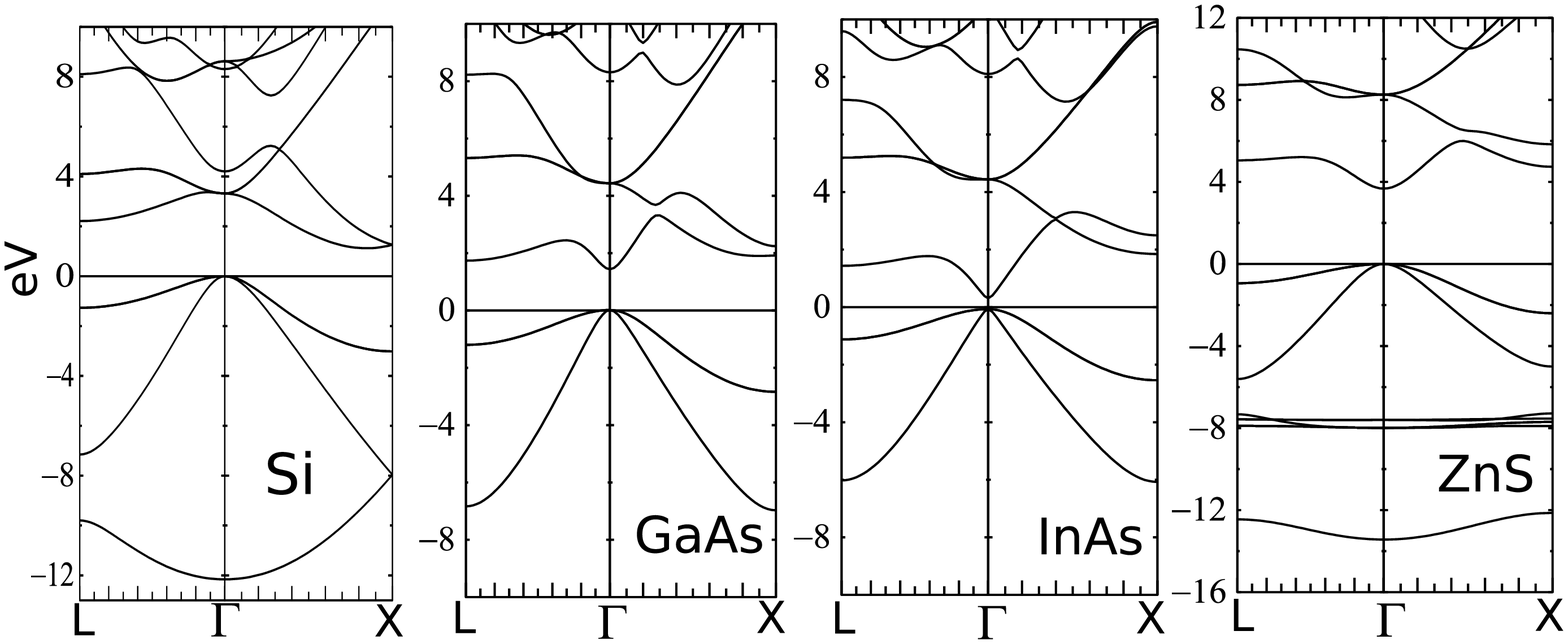}

\vspace{8mm}
\includegraphics[angle=0,scale=.4]{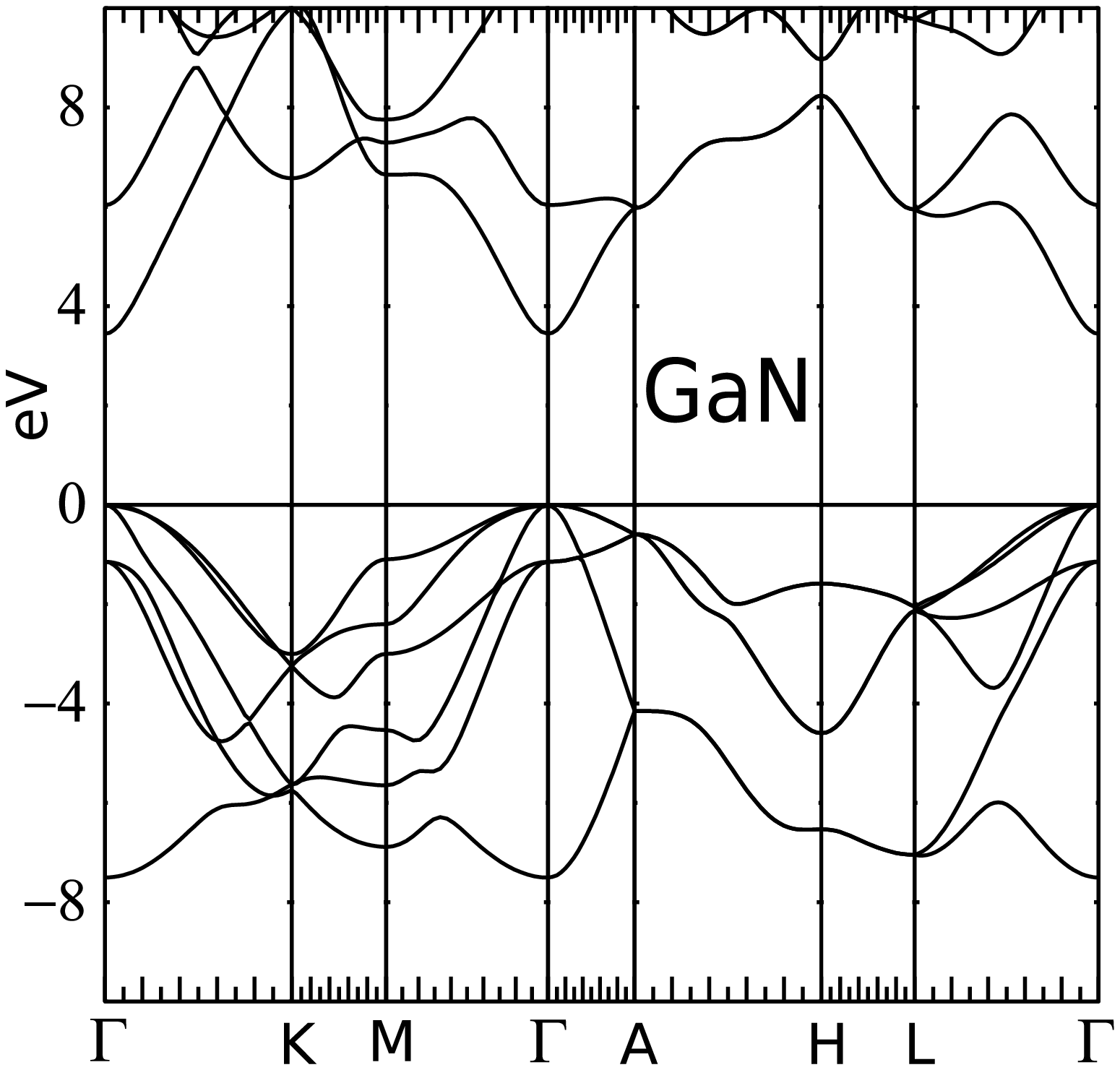}
\caption[]{Energy bands calculated by the \qsgw$\alpha$ 
method (\req{eq:alpha}), with 
$\alpha$ chosen to reproduce the experimental
band gap $E_{G}$. In wurtzite GaN, $E_{G}$=3.44eV~\cite{madelung96},
and $\alpha=0.79$.
Data for other compounds can be found in Table~\ref{tab:gap}.}
\label{fig:band}
\end{figure*}

\begin{table}
\caption{
\baselineskip 12pt
Eigenvalues of semiconductors relative to the valence band maximum 
at $\Gamma$. The QS$GW$ column depict results without spin-orbit coupling;
values in parentheses include spin-orbit coupling. 
Column QS$GW\alpha$  shows values after scaling defined in \req{eq:alpha}.
$\alpha$ is chosen so that the QS$GW\alpha$ potential (without spin-orbit
coupling) reproduces the experimental minimum band gap at room temperature.}
\begin{tabular}{l|*3{@{\hspace*{0.5em}}l}}
\colrule 
  Si       &  Expt.\ftn[1] & QS$GW$ & QS$GW\alpha$\\
           &                &        & $\alpha=0.85$ \\
  \ \  \ $\Gamma_{15c}$ &  3.34 & 3.45(3.41) & 3.32  \\
  \ \  \ $L_{6c}$       &  2.04 & 2.35 & 2.21  \\
  \ \  \ $E_g$          &  1.12 & 1.23(1.11) & 1.12  \\
  \ \  \ $\Gamma_{2'c}$ &  4.15   & 4.38 & 4.21  \\
\colrule 
  GaAs  &  Expt.\ftn[1] & QS$GW$ & QS$GW\alpha$\\
        &               &        & $\alpha=0.68$ \\
  \ \  \ $\Gamma_{6c}$  &  1.42 & 1.93(1.81) & 1.42  \\
  \ \  \ $L_{6c}$       &  1.66 & 2.11 & 1.72  \\
  \ \  \ $X_{6c}$       &  1.97 & 2.12 & 1.90  \\
  \ \  \ $\Gamma_{7c}$  &  4.50 & 4.74 & 4.42  \\
\colrule 
  InAs &                 empPP\ftn[2] & QS$GW$ & QS$GW\alpha$\\
       &                &        & $\alpha=0.65$ \\
  \ \  \ $\Gamma_{6c}$  &  0.37  & 0.79(0.68) & 0.38  \\
  \ \  \ $L_{6c}$       &  1.53 & 1.86 & 1.51  \\
  \ \  \ $X_{6c}$       &  2.28 & 2.10 & 1.90  \\
  \ \  \ $\Gamma_{7c}$  &  4.39  & 4.84 & 4.51 \\
\colrule 
  ZnS  & Expt. \ftn[1] & QS$GW$ & QS$GW\alpha$\\
       &               &      & $\alpha=0.83$ \\
  \ \  \ $\Gamma_{6c}$  &  3.68 & 4.04(4.01) & 3.68  \\
  \ \  \ $L_{6c}$        &      & 5.45 & 5.05  \\
  \ \  \ $X_{6c}$        &      & 5.05 & 4.74  \\
  \ \  \ $\Gamma_{7c}$  &      & 8.67 & 8.26 \\
\colrule
\end{tabular}
\footnotetext[1]{Experimental data at room temperature,
taken from Ref.~\ocite{madelung96}.}
\footnotetext[3]{Empirical pseudopotential
data are taken from Ref.~\ocite{madelung96}.
The experimental direct gap is $\sim$0.4eV.}
\label{tab:gap}
\end{table}


Here we treat Si, GaAs, InAs, zincblende ZnS, and wurtzite GaN. 
For each material we calculate a self-consistent noninteracting hamiltonian 
$\H0$ through the \qsgw\ formalism.
Spin-orbit coupling is neglected, following prior work \cite{sano95,jung96}.
Table \ref{tab:gap} shows calculated values at high-symmetry points, 
compared with available experimental data.
As we and others have noted\cite{kotani07a,vans06,shishkin07}, the \qsgw\ gap
is systematically overestimated because the RPA underestimates the screening.
(Also the GaAs calculation used a smaller basis what was reported 
in \cite{vans06}, resulting in 
an additional overestimate of $\sim$0.05 eV.)
To compare the \qsgw\ results to experiment, we must take into account
other contributions: spin-orbit coupling, zero-point motion\cite{cardona05},
and finite temperature all reduce the gap slightly\cite{mark06adeq}.

To obtain the most reliable IIR, we slightly modify $\H0$ to
reproduce the experimental gap at room temperature without
including these contributions explicitly.  To do this, we add an
empirical scaling (``\qsgw$\alpha$'' correction) following the
procedure used in Ref.\cite{chantis06a}.  We scale the one-body
Hamiltonian as follows:
\begin{eqnarray}
H^\alpha = \H0 + (1-\alpha) (\widetilde{\Sigma} -V_{\rm xc}^{\rm LDA})
\label{eq:alpha}
\end{eqnarray}
where $\widetilde{\Sigma}$ is the static version of the self-energy 
(see Eq.(10) in Ref.\cite{kotani07a}). 
Table \ref{tab:gap} shows numerical values both with and without
the scaling.  As we showed in Ref.\cite{chantis06a},
effective masses are also well reproduced.
Thus we can set up a satisfactory $H^\alpha$ with a single parameter
$\alpha$.  This procedure is reasonable because the uncorrected
gaps are already close to experiment and $1-\alpha$ is not large.
The materials-dependence of $\alpha$ shown in Table \ref{tab:gap}
originates largely from the dependence of SO coupling
and finite temperature on material.  If these were taken into
account by improving $\H0$ explicitly, a universal choice of
$\alpha\sim0.8$ would reproduce the experimental gaps in the
Table to within $\sim$0.1eV. (Alternatively, adopting the
present procedure with a universal $\alpha\sim0.75$ accomplishes much
the same thing.)  Table \ref{tab:gap} shows that the experimental energy
dispersions are also well reproduced where they
are well known (Si and GaAs).  This systematic tendency is found for many other
materials, including ZnO, Cu$_{2}$O, NiO and
MnO~\cite{Faleev04,kotani07a}, and GdN\cite{chantis07f}.  It
implies that the \qsgw$\alpha$ procedure is broadly applicable
with comparable accuracy to many environments, e.g, to InAs/GaAs 
grain boundaries.
The \qsgw$\alpha$ energy bands are shown in Fig.\ref{fig:band}.

\begin{figure*}[htbp]
\centering
\includegraphics[angle=0,scale=.6]{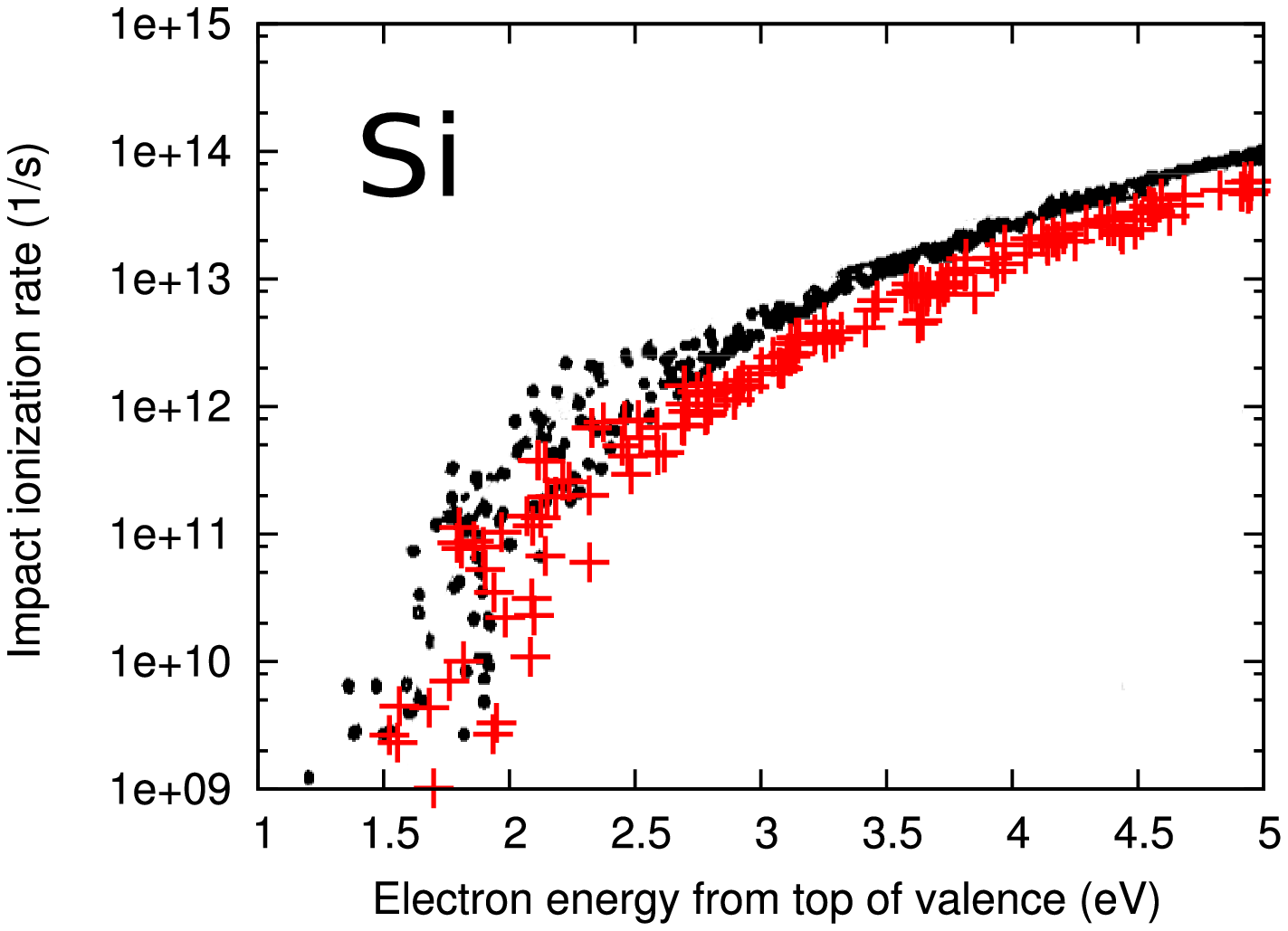}
\includegraphics[angle=0,scale=.6]{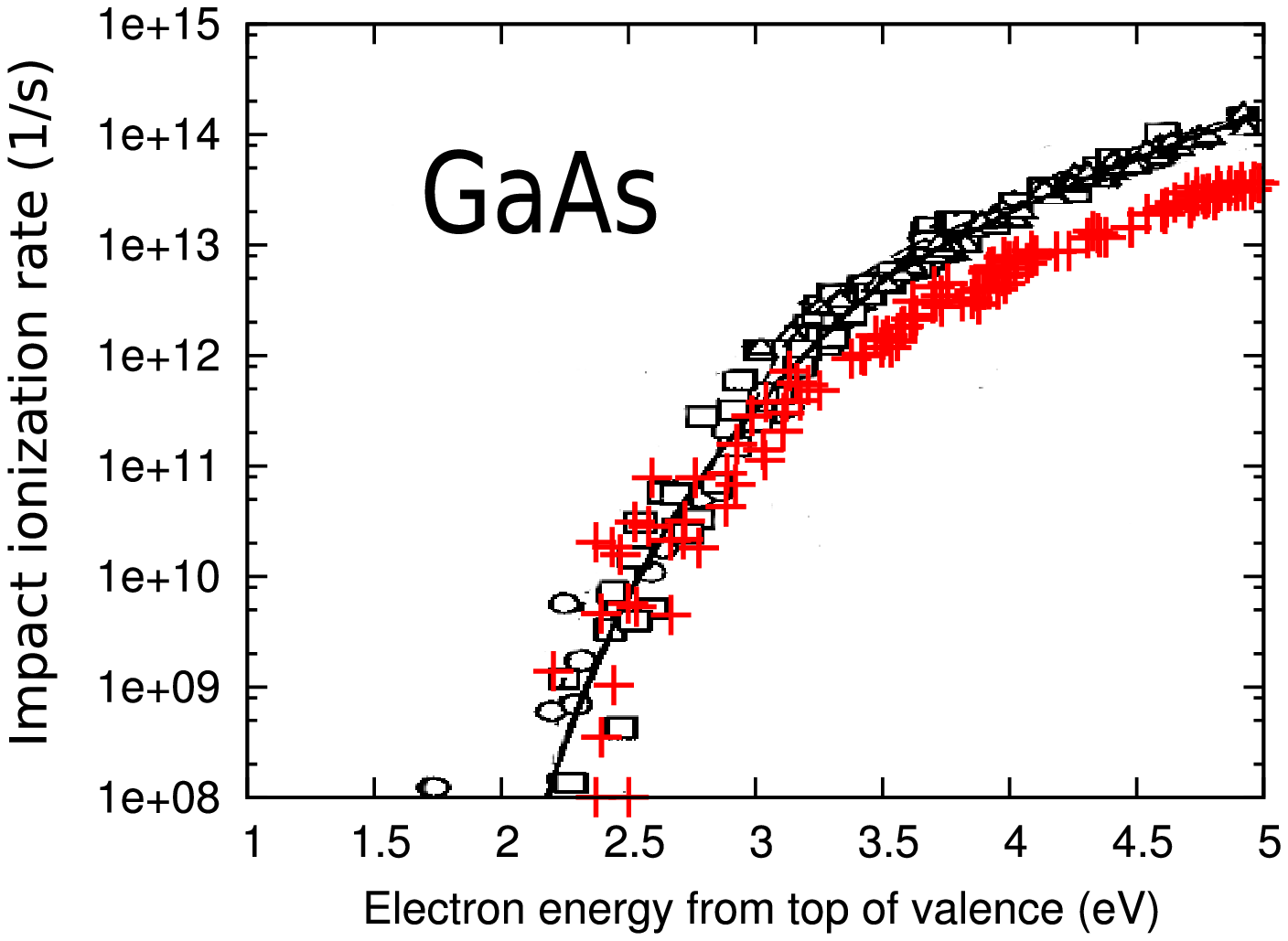}
\includegraphics[angle=0,scale=.6]{inasimi.epsi}
\includegraphics[angle=0,scale=.6]{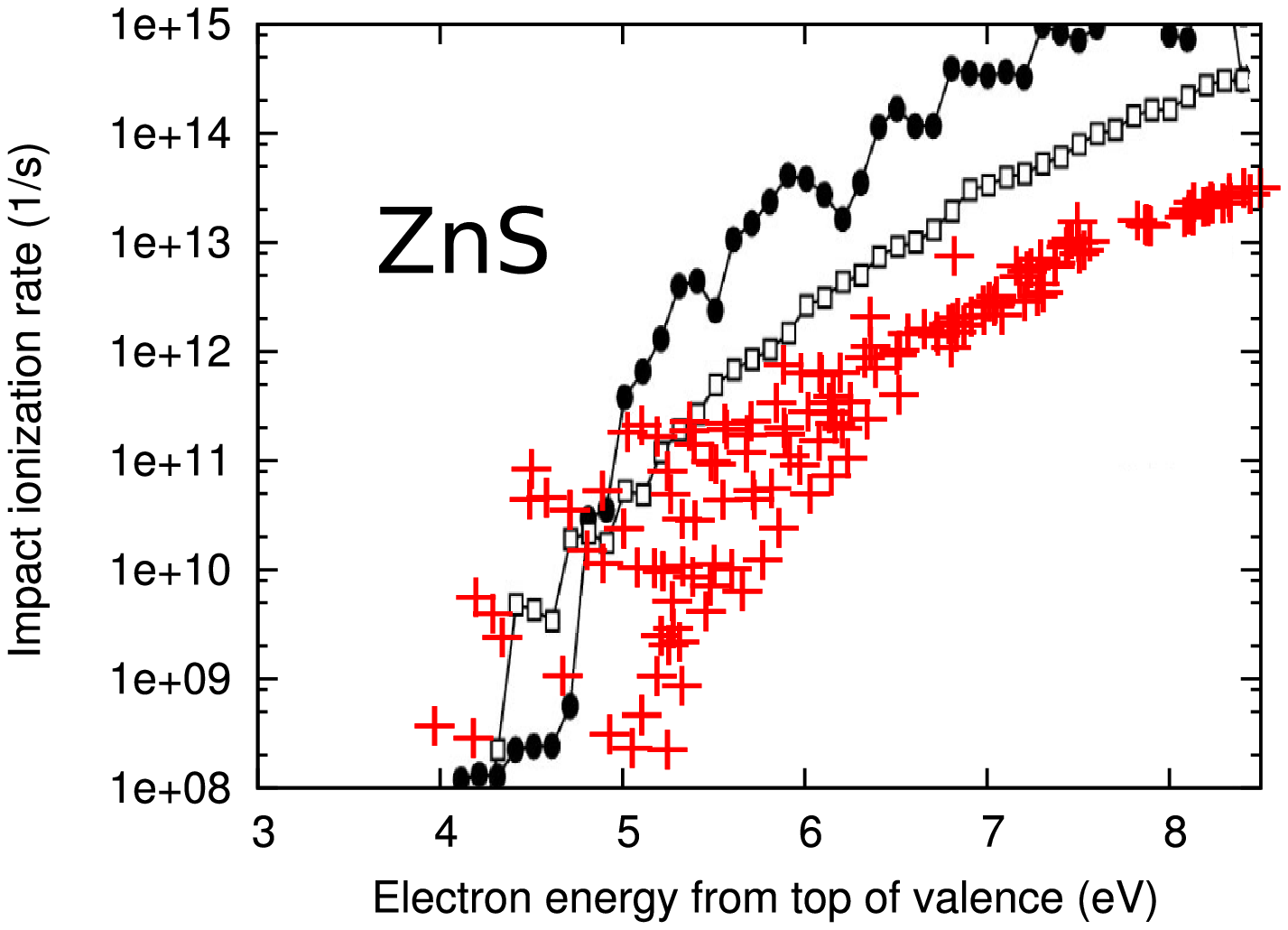}
\includegraphics[angle=0,scale=.6]{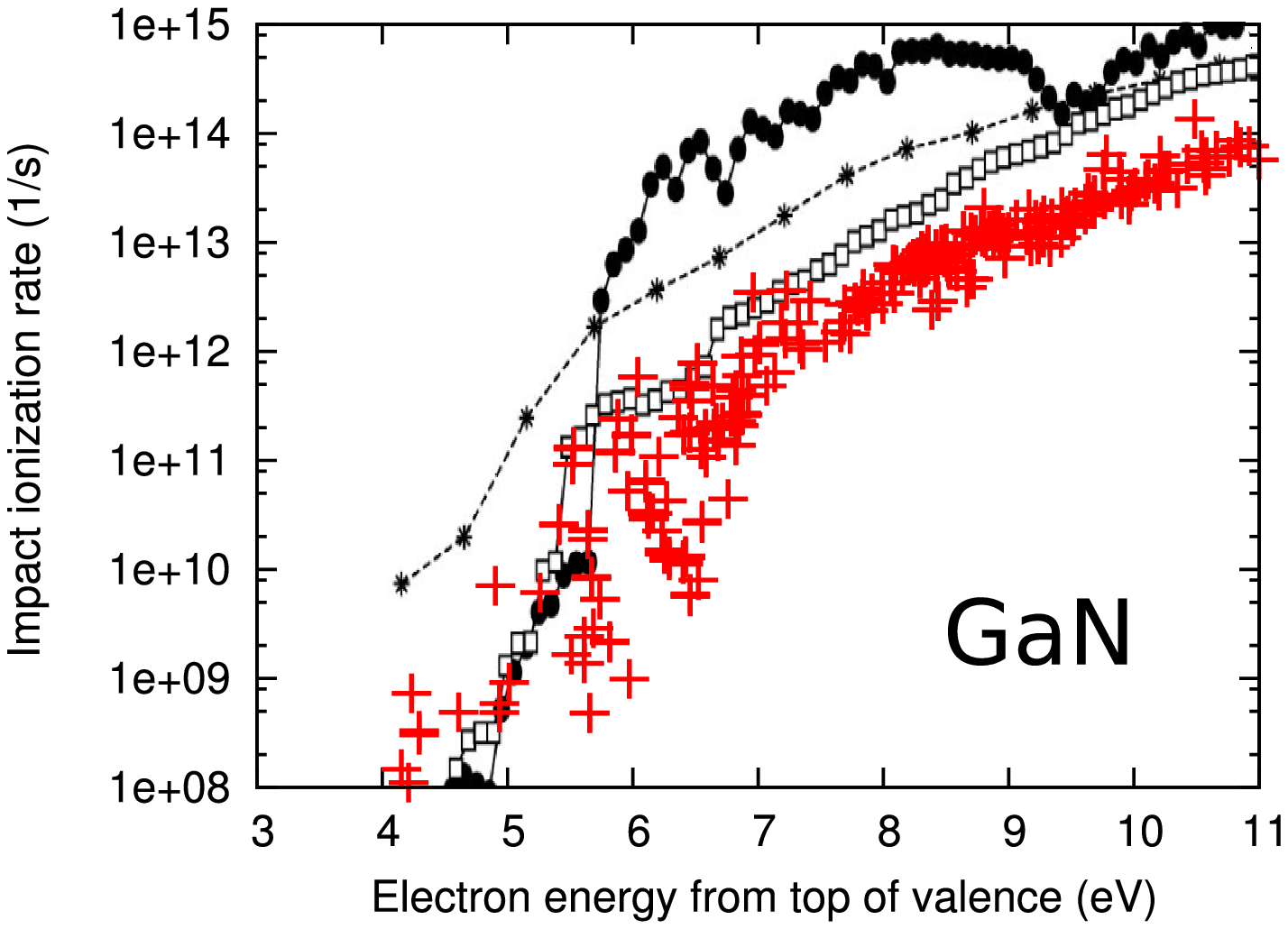}
\caption[]{(color online) Impact ionization rates 
$\tau_{\bfk n}^{-1}$
as a function of the initial electron energy $\varepsilon_{\rm in}$ 
(measured from the bottom of conduction band).  The present
$GW$A calculation is shown by large (red) plus signs.  It is superposed on
previous calculations: Si and InAs from Sano and Yoshii \cite{sano95},
GaAs from Jung et al. \cite{jung96}, ZnS and GaN from Kuligk et
al. \cite{kuligk05}.  Open boxes in the ZnS and GaN data are
EPP results from Picozzi et al.~\cite{picozzi02}; solid circles are 
exact exchange results~\cite{kuligk05}.
We used 500 $\bfk$ points in the 1st Brillouin zone for GaN, and
1728 points for the cubic compounds (regular mesh including the
$\Gamma$ point\cite{kotani07a}). Owing to the limited number of $\bfk$ points,
there are some numerical errors, e.g. a factor of order 2
when IIR $\lesssim$ 1e+10).  The error is not large enough
to affect our conclusions.}
\label{fig:imi}
\end{figure*}

Given $H^{\alpha}$, we perform a one-shot $GW$A 
calculation using the method detailed in Ref.\cite{kotani07a},
and calculate $\tau_{\bfk n}^{-1}$ from \req{eq:invt}.
To reduce the computational time we truncate
the product basis for each atomic site  
to $l\le{}1$.  This limits the degrees of freedom 
for the local-field correction in the dielectric function. 
However, we checked that this little affects the results.
To obtain $Z_{\bfk n}$ in \req{eq:invt}, 
we need to calculate the derivative of the 
self-energy $\frac{\partial {\rm Re} \Sigma(\omega)}{\partial \omega}$ 
at $\varepsilon_{\bfk n}$, though $Z$ contributes a relatively unimportant
factor $\sim 0.8$. 
The main computational cost of the IIR calculation comes from the sum 
of the pole weights on the real axis; see Eq.~(58) 
in Ref.\cite{kotani07a}. 
This corresponds to the convolution 
of Im$G$ and Im$W$, \req{eq:imsig},
after Im$W$ is obtained from integration by the tetrahedron method \cite{kotani07a}.
Fig.\ref{fig:imi} shows our results for $\tau_{\bfk n}^{-1}$. 
The $x$-axis denotes the initial electron energy 
$\varepsilon_{\rm in}$ measured from the bottom of the conduction band;
$\varepsilon_{\rm in}>E_G$ is a hard threshold below which IIR is zero.
The present results, depicted by large plus signs, are 
superposed on results taken from previous work.

The IIR has a typical feature 
as already shown in Fig.1 in Ref.\cite{sano94}, that is,
the IIR as function of $\varepsilon_{\rm in}$
are widely scattered at low $\varepsilon_{\rm in}$ 
because of the limited number of
transitions that conserve energy and momentum.
The scatter diminishes at high energy because 
of an averaging effect which smears  
the anisotropy in the Brillouin zone as discussed in \cite{sano94}. 
Our results for Si and GaAs correspond rather well to previous work.
Details for the IIR are already well analyzed \cite{sano92,sano94,sano95,jung96,kuligk05,picozzi02,picozzi02L}.

Turning to ZnS and GaN, we superpose our results
on those presented by Kuligk et al. in Figs. 9 and 10 of Ref.\cite{kuligk05},
which include exact exchange (EXX) (solid symbols) and EPP results
from Ref.\cite{picozzi02} (open symbols). The EPP and the present calculations
appear mostly similar apart from an approximately
constant factor; however the EXX results show rather different
behavior, particularly in GaN.  This is likely because
the EPP and \qsgw$\alpha$ energy bands are quite similar to each other,
but they are quite different from the EXX case (see Figs 2 and 3 in Ref.\cite{kuligk05}).


A large discrepancy with EPP is seen only in InAs. Our data is
superposed on the calculations by Sano and Yoshii \cite{sano95}. 
We obtain high IIR at low initial electron energies 
$\varepsilon_{\rm in}\gtrsim$1 eV. 
Such high IIR comes from 
initial electrons near the conduction band minimum at the $\Gamma$ point.
Since the band gap and effective mass are small in InAs, there
are states not far from $\Gamma$ with energy $\varepsilon_{\rm in}>E_G$,
which can generate an electron-hole pair.
This occurs only for InAs in the cases studied,
but generally occurs for narrow gap semiconductors.
For the discrepancy with results of
Sano and Yoshii may be due to their constant matrix elements approximation,
which is not suitable for such a narrow gap material
(see Ref.\cite{sano95} near Eq.(2)).
%

In conclusion, we have calculated the IIR for several
materials in the $GW$A, after a theoretical 
discussion of its application to the IIR.
In principle, the method presented here 
will be applicable even to inhomogeneous systems
such grain boundaries and quantum dots where we expect very strong IIR.
The present calculations correspond reasonably well to prior work,
with the exception of the narrow gap material InAs.
High IIR would be expected universally
in similar narrow gap materials such as GaSb, InSb, and
InN.  This indicates that careful consideration
for the IIR might be required when we use such materials 
for devices. 

This work was supported by ONR contract N00014-7-1-0479,
and the NSF QMHP-0802216.
We are also indebted to the Ira A. Fulton High Performance 
Computing Initiative.

\bibliography{imi,lmto,gw}

\end{document}